\documentclass[twocolumn,showpacs,preprintnumbers,amsmath,amssymb,aps,prl,lengthcheck]{revtex4-1}
\usepackage{amssymb,color,array}
\usepackage{graphicx}
\usepackage{amsmath}
\usepackage{hyperref}

\begin{document}

\title{Weak measurement protocols for Majorana bound state identification}

\author{J. Manousakis,$^{1,2}$ C. Wille,$^3$ A. Altland,$^1$ R. Egger,$^4$ K. Flensberg,$^2$ and F. Hassler$^5$} 
\affiliation{${}^1$~Institut f\"ur theoretische Physik, Universit\"at zu K\"oln, Z\"ulpicher Stra{\ss}e 77, D-50937 K\"oln, Germany\\
${}^2$~Center for Quantum Devices, Niels Bohr Institute, University of Copenhagen, DK-2100 Copenhagen, Denmark\\
${}^3$~Dahlem Center for Complex Quantum Systems, Physics Department, Freie Universit\"at Berlin, D-14195 Berlin, Germany\\
${}^4$~Institut f\"ur Theoretische Physik, Heinrich Heine Universit\"at, D-40225 D\"usseldorf, Germany\\
${}^5$~JARA-Institute for Quantum Information, RWTH Aachen University, D-52056 Aachen, Germany}

\date{January 2020}

\begin{abstract}
We propose a continuous weak measurement protocol testing the nonlocality of Majorana bound states through current shot noise correlations. The experimental setup contains a topological superconductor island with three normal-conducting leads weakly coupled to different Majorana states.  Putting one lead at finite voltage and measuring the shot noise correlations between the other two (grounded) leads,  devices with true Majorana states are
distinguished from those without by strong current correlations. The presence of true Majoranas manifests itself in unusually high noise levels or the near absence of noise, depending on the chosen device configuration. Monitoring the noise statistics amounts to a weak continuous measurement of the Majorana qubit and yields information similar to that of a full braiding protocol, but at much lower experimental effort.
Our theory can be adapted to different platforms and should allow for clear identification of Majorana states.
\end{abstract}

\maketitle

\emph{Introduction.---} Throughout the past decade the quest for stable realizations
of Majorana bound states (MBS) has become a major theme in condensed matter
physics\cite{Kitaev2001,Nayak2008,Alicea2012,Leijnse2012,Beenakker2013,Sarma2015,Aguado2017,Lutchyn2018}.
A fully manipulable MBS would pave the way to disruptive
developments, both in fundamental science, and as a building block for a new
generation of quantum
hardware\cite{Alicea2011,Terhal2012,Vijay2015,Landau2016,Plugge2016,Plugge2017,Karzig2017,Litinski2017,Wille2019}.
While initial proposals were focusing on realizations as end states in topological
semiconductor quantum wires, the quest for the Majorana has led to the recent
discovery of various alternative material platforms
\cite{Liu2018b,Zhang2018b,Wang2018b,Sajadi2018,Ghatak2018,Murani2019}. In all these,
evidence for Majorana states has been reported on the basis of tunneling spectroscopy
or related \emph{local} probes, see, e.g.,
Refs.~\cite{Lutchyn2018,Mourik2012,Yazdani2014,Ruby2015,Albrecht2016,Deng2016,Nichele2017,Suominen2017,Gazi2017,Zhang2018,Ren2019,Vaiti2018,Deacon2017,Bocquillon2017,Laroche2018,Fornieri2018}.
However, in spite of promising signatures, more mundane explanations, such as Andreev bound states representing pairs of `fake' Majorana states, cannot be ruled out, and the interpretation of the experiments remains  debated,
  cf.~Refs.~\cite{Altland2012,Liu2012,Chiu2018,Zazunov2018,Liu2018,Moore2018,Vuik2018,Zhang2019a,Cayao2015,San-Jose2016,Reeg2018,Prada2012,Kells2012,Stanescu2013,Fleckenstein2018,Liu2017,Moore2018b,Stanescu2019,Penaranda2018,Avila2019}. In view of this situation, various forms of diagnostics transcending tunneling
spectroscopy have been
proposed \cite{Lai2019,Tuovinen2019,Smirnov2018,Cornfeld2019,Jonckheere2019,Hell2018,Rubbert2016,Fu2010,Schrade2019,Haim2015,Haim2015b,Liu2015,Tripathi2016,Jonckheere2017,Grabsch2019,Beri2012,Altland2013,Beri2013,Gau2018,Sela2019,Guerci2019,Hansen2018,OFarrell2018,Danon2019,Deng2018,Prada2017,Cayao2017,Zazunov2012,Virtanen2013,Ioselevich2016,Zazunov2017,Zazunov2018b,Cayao2018a,Schuray2018,Schrade2018,Cayao2018b,Awoga2019,Schulenborg2019}.
 Basically, these fall into two categories, local probes corroborating evidence for
the presence of genuine Majoranas albeit still containing potential loopholes, or compelling probes, such as braiding protocols, which, however, do
not seem to be a realistic option in the immediate future. 

In this Letter, we suggest a new type of diagnostic experiment. The strategy will be  to
access the  information stored nonlocally in a set of at least three MBS through the \emph{statistical fluctuations} of tunneling
current probes. As we are going to show below, this yields information comparable to that of  a full fledged
braiding protocol, but at much lower experimental effort. In fact, the
hardware  required to perform the measurement is not much different from that currently in operation and should be realizable for the  proposed
Majorana platforms by present-day technology.
We note that statistical fluctuations of tunneling
current probes have also been investigated in other studies
of topological systems,  see, e.g., Refs.~\cite{Chevallier2016,Mi2018,Souto2019}.

\begin{figure}[t] 
 \includegraphics[width=\columnwidth]{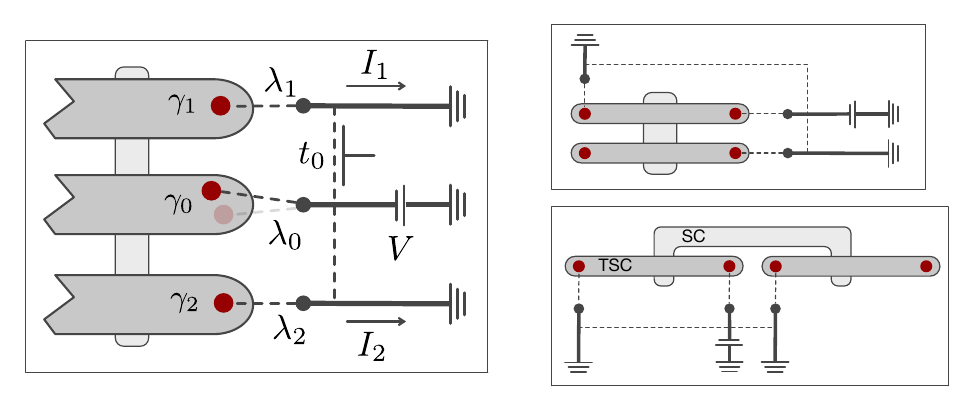}
\caption{Setup for probing Majorana bound states in a system of topological
quantum wires. Left: Three of the Majorana states (dots) on a Majorana-Cooper box with three topological hybrid nanowires connected by a superconducting backbone \cite{Plugge2017,Karzig2017} are tunnel coupled to  normal-conducting leads. The schematic on the left indicates that one of the leads ($\alpha=0$) is biased with a voltage $V$ and acts as a source of electrons into the  grounded drain leads ($\alpha=1,2$). Tunable tunnel couplings $t_0$ introduce a direct link between the source and drain leads. 
 Andreev states are distinguished from genuine Majoranas as pairs of MBS $\gamma_\alpha^i$ (with $i=1,2)$ centered close to the tunnel interface (cf. faded dot, representing an $i=2$ state in the  wire $0$).
 The cross-correlation shot noise amplitude $S_{12}$ of the 
 currents $I_1$ and $I_2$ (Eq.~\eqref{s12}) unambiguously distinguishes between the two types of states. Right: The same experiment can be carried out on a wide range of possible device layouts.  The two schematics on the right give examples for additional realistic geometries
 using only two nanowires.}
\label{fig1} 
\end{figure}

Before turning to a more detailed discussion, let us sketch the idea of the approach.
Consider the schematic representation of Fig.~\ref{fig1}, where the dots represent
MBS supported on a floating mesoscopic superconductor (see the right panels
for more realistic layouts). Suppose we measure the tunneling current, $I_1(t)$,
flowing in response to a  voltage bias applied at the wire connecting to MBS
$\gamma_1$ relative to a grounded wire connecting to $\gamma_0$. This current is
sensitive to the state of the qubit operator $\sigma_1\equiv i \gamma_1
\gamma_0$\cite{Plugge2016,Plugge2017}.  Monitoring the current over short intervals
of time, a  weak  measurement\cite{Clerk2010,Wei2008}  is effectively performed, continuously steering the qubit
into a state defined by the current readout. Now assume that the current, $I_2(t)$, through
terminal $2$ is recorded as well. This readout couples to $\sigma_2\equiv i
\gamma_2 \gamma_0$ and the tendency to alter this operator, non-commuting with
$\sigma_1$, implies incompatible readouts. Its observable consequence is  pronounced
current cross-correlations, which we will demonstrate represent a unique signature in
that they are qualitatively distinct from the noisy current in the presence of
Andreev bound states, or other low energy quasiparticle (poisoning) excitations. More specifically, our prime observable of interest is the current cross correlation,
\begin{equation} \label{s12}
    S_{12}=\int dt \ \langle\langle I_1(t) I_2(0) \rangle\rangle  ,
\end{equation}
where $\langle \langle A B \rangle \rangle = \langle A B\rangle-\langle A\rangle\langle B\rangle$. We will analyze this quantity both in the presence and absence of tunneling bridges (cf. vertical dashed lines in Fig.~\ref{fig1}) between the electrodes connecting to the island. This additional structure, which can be controlled during an experiment via gate electrodes, gives us sufficient information to distinguish MBS from the competing cases mentioned above. This is because the noise profile  probes the presence of an underlying Pauli algebra, which is a unique characteristic of the Majorana system (alternatively diagnosed in a more elaborate braiding protocol).

\emph{Model.---} We describe the setup of Fig.~\ref{fig1} by the now standard \cite{Fu2010,Beri2012,Altland2013,Beri2013} Hamiltonian $H= H_C + H_\text{leads} + H_T + H_\text{ref}$ for a `Majorana-Cooper box'. Here 
 $H_C = E_C (N - n_g)^2$ defines the charging energy $E_C = 2e^2/C$ associated with $N = - i \partial_\phi$ Cooper pairs 
on the floating island ($\phi$ is the phase of the superconductor). 
We consider Coulomb valley conditions defined by a backgate parameter $n_g$  close to an integer value.
The normal-conducting leads, $\alpha=0,1,2$, are modeled by a standard noninteracting Hamiltonian, $H_{\rm leads}$,
with electron annihilation operators $c_{\alpha,k}$ for momentum $k$ and  
 density of states, $\nu_\alpha=\nu$, assumed equal for simplicity.
The local tunneling  between the Majorana box and the leads is described by
\begin{equation}\label{HT}
H_T = \sum_{\alpha=0,1,2}  \sum_{j=1}^{N_\alpha} \lambda_\alpha^j c^\dag_{\alpha} \gamma_\alpha^j e^{-i\phi/2} + \text{h.c.},
\end{equation}
where $c_\alpha=\sum_kc_{\alpha,k}$. Here, $\gamma_\alpha^j$ represent the low energy
box states at terminal $\alpha$. Representing them by the Majorana operators
$\gamma_\alpha^j=\left(\gamma_\alpha^j\right)^\dagger$,
$\{\gamma_\alpha^j,\gamma_{\alpha'}^{j'}\}=2\delta_{\alpha\alpha'}\delta_{jj'}$,
tunnel coupled by amplitudes $\lambda_\alpha^j$, this modeling includes the cases
$N_\alpha=1$  of a genuine Majorana, and   $N_\alpha>1$ where Andreev states
described as pairs of spatially overlapping Majoranas\cite{Schrade2019}  compromise
the system. Taking note that Majorana states carry no charge, the operator
$e^{-i\phi/2}$  in Eq.~\eqref{HT} accounts for the removal of an island electron
charge upon tunneling.  Finally, the reference arms in Fig.~\ref{fig1} are modeled by
$H_\text{ref} = \sum_{\alpha=1,2} t_{0,\alpha} c_{\alpha}^\dag c_{0}^{} + \text{h.c.}$, with
the gate-tunable tunneling amplitude $t_{0,\alpha}$. With $V$ denoting the voltage bias applied to the source lead ($\alpha =0$) and $\Delta$ the superconducting gap on
the island, we consider the parameter regime ($e=k_B=\hbar=1$ throughout)  $|\lambda|,V\ll E_C,\Delta$, at low
temperatures, $T \ll V$.  In this case, transport through the island
is dominated by cotunneling processes and second-order perturbation theory in the
$\lambda_\alpha^j$ yields the effective Hamiltonian $H_C+H_T\to \tilde H_T$ with
\begin{eqnarray}
    \tilde H_T &=& \frac12 \sum_{\alpha\neq \alpha'} \mathcal{O}_{\alpha\alpha'} c^\dag_{\alpha} c_{\alpha'}^{} 
    + \text{h.c.} , \nonumber \\
     \label{oaa}
    \mathcal{O}_{\alpha\alpha'} &=& i\sum_{j=1}^{N_\alpha} \sum_{j'=1}^{N_{\alpha'}} t^{jj'}_{1,\alpha\alpha'}  \gamma_\alpha^j \gamma_{\alpha'}^{j'},
\end{eqnarray}
where $t^{jj'}_{1,\alpha\alpha'} \simeq i \lambda^j_\alpha
(\lambda^{j'}_{\alpha'})^\ast_{}/E_C$. Where possible, we use the simplified notation
$t_{1,\alpha\alpha'}^{jj'}= t_1$ and $t_{0,\alpha}=t_0$ throughout. The results discussed below are all perturbative to leading order in the dimensionless tunnel conductances $g_0 \equiv 2\pi \nu^2 |t_0|^2$ and $g_1 \equiv 2\pi \nu^2 |t_1|^2$ characterizing the different connectors between leads. We assume these to be tuned to $g_0 \ll 1$ and $g_1 \ll 1$, conditions that can be checked by designated calibrating measurements.

\emph{Qualitative discussion---} If the wires host single Majorana states,
$N_\alpha=1$, the projection to the quantized charge sector implies the parity
constraint $\gamma_0\gamma_1\gamma_2\gamma_3=\pm 1$, where the presence of the
disconnected Majorana $\gamma_3$   is required to define a complete system of
wire-end states \cite{Beri2012}.  The ground state then is doubly degenerate and
defines a qubit with the above Pauli operators $\sigma_{1,2}$ \cite{Plugge2017,Karzig2017}. To lowest order in
perturbation theory in the tunneling amplitudes, the average currents flowing through
the drain leads are given by
\begin{align}\label{eq:i_mbs_stat_dep}
     \langle I_{\alpha=1,2} \rangle = \left(g_0+g_1+ 2g_\mathrm{i} \langle \sigma_\alpha\rangle \right) V  ,
\end{align}
where the interference factor $g_\mathrm{i}\equiv 2\pi \nu^2 \operatorname{Re}(t_0^* t_1)$ couples to the
measured eigenvalue of the respective Pauli operator $\sigma_\alpha$
\cite{Fu2010,Landau2016,Plugge2016}. In a way made more rigorous below, the recording
of the \emph{simultaneously flowing} currents $I_\alpha$ in
a measurement of the cross correlation $S_{12}$ amounts to a continuous weak measurement of $\sigma_1$ and
$\sigma_2$. This view implies that the system cannot settle
in a pure state, because such a state would need to be a simultaneous eigenstate of
$\sigma_1$ and $\sigma_2$. The observable consequence of this frustration are
persistent  fluctuations of $I_1$ and $I_2$, quantified by $S_{12}$, Eq.~\eqref{s12}.
Below we will show how this principle implies a positive cross correlation
$S_{12}\sim F \bar I $, where $\bar I$ is the average current, and $F$ a Fano factor of $\mathcal{O}(1)$. As will be discussed below, this should be compared to  parametrically smaller results proportional to higher powers of the tunneling conductance characterizing noninteracting electrons in the tunneling limit
\cite{Martin1992}. The origin of stronger correlations in the present system is the coupling of transport to a  Pauli algebra 
 which effectively conditions the currents in the arms $1$ and $2$ to each other. 

\emph{Counting statistics.---} We next derive an efficient formalism to compute the
cross correlation $S_{12}$, and related statistical signatures of transport. The
first step is to integrate over the lead degrees of freedom to obtain a reduced
density matrix $\rho_t$ in the Hilbert space corresponding to the Majorana operators $\gamma_{\alpha}^j$. While this object by itself is not too informative, the
statistics of the charge $Q_\alpha$ transmitted in time $t\in [\tau_0,\tau_0+\tau]$ through the terminals
$\alpha=1,2$ is obtained by introducing  counting field factors $e^{\pm
if(t)\chi_\alpha/2}  $ into the hopping amplitudes $\lambda_\alpha^j$ and
$t_0$, where $f(t)=1$ in the time interval of observation and  zero otherwise,
the quantities $\chi_{1,2}$ are
constant counting fields, and the sign 
factor refers to counting fields on the forward or backward time evolution in
$\rho_t= e^{-iH t} \rho_0 e^{iHt}$. 
 Defining $z_\alpha=\exp(i\chi_\alpha)$ the density
matrix, $\rho_{\tau}(z_1,z_2)$, then depends on the counting parameters $z_\alpha$,
and all cumulants of the charges $Q_{1,2}$  are obtained by taking derivatives\cite{Wei2008,Franquet2017,Franquet2017b}, 
\begin{equation}
\langle \langle Q_1^n Q_2^m \rangle \rangle =  (z_1\partial_{z_1})^n (z_2\partial_{z_2})^m\big|_{z=1}\ln \operatorname{Tr}\rho_{\tau}(z_1,z_2).
\end{equation}
The evolution equation governing $\rho_t=\rho_t(z_1,z_2)$,
is given by~\cite{appendix}
\begin{eqnarray}
  \label{eq:BR}
     \dot \rho_t &=& -i \left[H_q, \rho_t\right]  + 2\pi \nu^2 T \left( \mathcal{D}_{12}(\rho_t) +  \mathcal{D}_{21}(\rho_t)  \right)  \\
    &+&2\pi \nu^2 V   \sum_{\alpha=1,2} \Bigl[   (z_\alpha -1) (\mathcal{O}_{\alpha 0}+t_{0,\alpha}) \rho_t
    ( \mathcal{O}_{\alpha 0}^\dag+t_{0,\alpha}^\ast)
    \nonumber \\  \nonumber
    && \qquad \quad - (z_\alpha -1)\mathcal{O}_{\alpha 0}\rho_t \mathcal{O}_{\alpha 0}^\dag + \mathcal{D}_{\alpha 0 }(\rho_t) 
    \Bigr].   
\end{eqnarray}
where the  superoperators 
\begin{equation}
    \mathcal{D}_{\alpha\alpha'}(\rho) = \frac{z_\alpha}{z_{\alpha'}} \mathcal{O}_{\alpha\alpha'} \rho \mathcal{O}_{\alpha\alpha'}^\dag -\tfrac12\left\{\mathcal{O}_{\alpha\alpha'}^\dag \mathcal{O}_{\alpha\alpha'}, \rho \right \} ,
\end{equation}
act as Lindbladians generalized for the counting parameters, $z_{1,2}$,
$z_0\equiv 1$, and $\mathcal{O}_{\alpha\alpha'}$ describes the  electron transfer  from lead $\alpha'\to \alpha$, see Eq.~\eqref{oaa}. The coherent evolution in Eq.~\eqref{eq:BR} is generated by the effective Hamiltonian 
\begin{eqnarray}\nonumber
    H_q &=& -\nu^2\Lambda \sum_{\alpha=1,2} \left( t_{0,\alpha}^\ast\mathcal{O}_{\alpha 0} +\text{h.c.} \right) \\ \nonumber &-& 
    \frac{\nu^2 \Lambda}{2} \sum_{\alpha<\alpha'}\left \{\mathcal{O}_{\alpha\alpha'}^\dag, 
    \mathcal{O}_{\alpha\alpha'} \right \}\\ & +&  \label{hq0}
    \nu^2 V \ln(\Lambda /2V) \sum_{\alpha=1,2} \left [\mathcal{O}_{\alpha 0} , \mathcal{O}_{\alpha 0}^\dag \right],
\end{eqnarray}
where $\Lambda \gg   V$ is the  bandwidth of the leads.  

\emph{True Majorana case }--- In spite of its complicated looking appearance,
Eq.~\eqref{eq:BR} can be solved, at least to the linear order in $V$ relevant to us.
We first note that in the absence of counting parameters, $z_1=z_2=1$, the stationary
solution approaches the isotropic limit $\rho_0=\frac{1}{2}\Bbb{I}_2$ at a time scale
$1/\Gamma$. The rate $\Gamma = 2g_1 V$ equals twice the average current flowing through the contact to MBS $\gamma_1$, indicating that the latter sets the time scale for the loss of information about the
initial states. 
Generalizing to the case of finite counting fields, we obtain~\cite{appendix} ($z= \tfrac12 z_1 +  \tfrac12 z_2-1$)
\begin{eqnarray}\label{eq:fcs}
  \ln\operatorname{Tr}  \rho_{\tau}(z_1,z_2) &=& \frac{\Gamma\tau}{2}\bigg(-1 +\frac{2 g_0+ g_1}{g_1}  z \nonumber\\
  &+& \sqrt{(1+z)^2 + \frac{8g_\mathrm{i}^2 z^2}{g_1^2}}\bigg).
\end{eqnarray}
This result yields the full counting statistics to order $V$. Specifically, the
stationary limit of the current, $\bar I =\langle I_\alpha \rangle$, through lead
$\alpha=1,2$ is given by $\bar I = \tau^{-1} \partial_{z_\alpha}
\ln\operatorname{Tr}\rho= (g_0+g_1) V$. This result is independent of $\sigma_\alpha$
and hence in stark contrast to Eq.~\eqref{eq:i_mbs_stat_dep}. It reflects the fact that the
continuous weak measurement of two non-commuting Pauli operators has eradicated
information about the qubit state and sent the system to a fully mixed state.
However, at the same time, one generically encounters an increased level of shot
noise cross correlations, $S_{12} = \tau^{-1}  \partial^2_{z_1,z_2}
\ln\operatorname{Tr}\rho =  F \bar I$. Here, $F$ is the positive Fano factor, which at $T=0$ and in the limit $\Gamma \tau \gg 1$
is obtained from Eq.~\eqref{eq:fcs} as
\begin{equation}\label{eq:Ftrue}
  F = \frac{2g_\mathrm{i}^2}{g_1 (g_0 + g_1 ) }.
\end{equation}
The most important message conveyed by this result is that
$F \sim 1$, parametrically exceeding $|F| \sim g_0$ in the non-interacting
limit\cite{Martin1992}. Also notice that the $\mathcal{O}(1)$ contribution to the
zero temperature Fano factor  vanishes identically  for pinched off reference
arms, $t_0\rightarrow 0$. This is because the continuous measurement of
$I_{\alpha}\propto |t_0 + t_1 \sigma_\alpha|^2$ no longer couples to two commuting
variables, $I_{\alpha,t_0\rightarrow0} \propto |t_1|^2$, and the
mechanism of large fluctuations no longer operates. For finite but low temperatures, thermal correlations produce a non-vanishing result for $t_0\to 0$ with, however, a very small Fano factor, $|F| \sim T/V \ll 1$.
This discussion shows how a comparison of cross correlations with and without
reference arms in one experimental setup will produce qualitatively different
results signifying the presence of a Pauli algebra. Furthermore, we note that small hybridizations of the MBS on the box do not affect this qualitative picture~\cite{footnote}.

\emph{Andreev bound states.---} We next discuss how the transport statistics change if at least one of the wires contains an Andreev bound state, $N_\alpha>1$. For definiteness, consider the case  $N_0=2$, $N_{1}=N_2=1$ without reference arms, where the source wire harbors an Andreev- instead of a Majorana state. We now need to differentiate between tunneling amplitudes, where $t_{1,\alpha\alpha'}^{jj'}$ for $\alpha=0$ and $j=1,2$ refers to the couplings between the source lead and the two MBS constituting the Andreev state. The resulting formulae of $S_{12}$ are more cumbersome. For example, under the simplifying assumption  $|t^{jj'}_{1,10}| = |t^{jj'}_{1,20}|$ and $\operatorname{Im}(t^{11*}_{1,10} t^{12}_{1,10})=-\operatorname{Im}(t^{11*}_{1,20} t^{12}_{1,20})$, we obtain~\cite{appendix}
\begin{equation}\label{sss12}
    F = \frac{[\operatorname{Im}(t^{11*}_{1,10} t^{12}_{1,10})]^2}{( |t^{11}_{1,10}|^2+|t^{12}_{1,10}|^2)^2}.
\end{equation}
Except for fine tuned choices, we always have $|F|\sim 1$, as in the true Majorana
case with reference arms. This high noise level again originates in the non-commuting nature of the 
 operators
${\cal O}_{10}$ and ${\cal O}_{20}$  (although they do not realize a Pauli algebra anymore).

\begin{table}
    \centering
    \begin{tabular}{|l|c|c|}\hline
         & $t_0 \neq 0$ & $t_0 = 0$ \\\hline
        true Majoranas ($N_0 = 1$, $N_{1,2}=1$) &  $\sim 1$ & $\ll 1$   \\
        Andreev bound states ($N_0 =2, N_{1,2} =1,2$) & $\sim 1$ & $\sim 1$ \\
        Andreev bound states ($N_0=1, N_{1,2} =2$) & $\sim 1$ & $\ll 1$ \\ \hline
    \end{tabular}
    \caption{Qualitative behavior of the Fano factor, $|F|$, for true vs fake Majorana states. The key observable distinguishing between the two cases is the large value $|F|$ in the absence of reference arms, $t_0=0$, for Andreev bound states. Since the protocol diagnoses only Andreev states coupled to the source lead, $\alpha=0$, experiments have to be repeated for different choices of the source and drain leads.}
    \label{tab:cross_corr}
\end{table}

With these results at hand, we propose a protocol to distinguish true vs fake
Majorana states, see Table~\ref{tab:cross_corr} for a summary: For true MBS and without reference
arms, $t_0=0$, the Fano factor $|F| =\mathcal{O}(T/V,g_{1,2})$ is parametrically smaller than the values $|F|=\mathcal{O}(1)$ predicted in their presence,  $t_0\not=0$. If the \emph{source terminal} is coupled to an Andreev bound state, strong cross-correlations with $|F|\sim 1$, regardless
of the presence or absence of reference arms are observed. This insensitivity of the noise level to the presence of the link clearly signals the presence of an Andreev state coupled to the source terminal. However, the protocol is blind to the presence of such states in the drain leads, cf. the third row of the table. It must therefore be repeated with the role of source and drain interchanged, which amounts to a different choice of bias voltages. On top of that, two more control measurements, must be performed, likewise by variation of gate or bias voltages:  (a) To exclude  false interpretations based on the measurement $|F|\ll 1$ due to the accidental fine tuning of parameters (e.g.,
$\operatorname{Im}(t^{11*}_{1,10} t^{12}_{1,10})$ in Eq.~\eqref{sss12}), the protocol
should be  repeated several times with  different values of the
gate potentials regulating the tunneling amplitudes. (b) We repeat that all results above hold to leading order in the tunnel conductances $g_\alpha$. To check for the presence of corrections in these parameters, one may repeat the protocol for a sequence of gradually diminishing conductances (adjustable by gate voltages). In the cases labeled $\sim 1$ in table~\ref{tab:cross_corr} this will leave the Fano factor parametrically unchanged, while for $\ll 1$ a suppression $\sim g_\alpha$ is predicted.

\emph{Quasiparticle poisoning ---} The transient in- and out-tunneling of
quasiparticles through MBS represents a source of decoherence and noise which, if
sufficiently strong, might compromise the interpretation of the zero frequency noise
correlators $S_{12}$. For completeness we therefore summarize a protocol~\cite{Plugge2017} geared to the characterization of quasiparticle poisoning processes. Consider both $t_{0,1}=\lambda_1^j=0$ such that lead 1 remains decoupled.
The current $I_2$ in Eq.~\eqref{eq:i_mbs_stat_dep} then depends on the state of the
MBS through the expectation value of $\sigma_2$ (or, more
generally, that of an operator $\mathcal{O}_{20}$ if Andreev bound states are
present). Beyond a time scale $\tau_\text{proj}
\simeq (g_0+g_1)/ (4g_\mathrm{i})^2 V$~\cite{appendix}, the measurement of $I_2$ becomes
projective, and  a weakly fluctuating result defined by one of the two values
$\langle \sigma_2\rangle\to \pm 1$ in Eq.~\eqref{eq:i_mbs_stat_dep} is approached. However, quasiparticle
tunneling accidentally switching the state $\sigma_2\to -\sigma_2$ will cause discrete jumps $\bar I_2\to\bar I_2\pm 4g_\mathrm{i}V$ in the readout. This should allow for a detection of quasiparticle induced decoherence.

\emph{Large fluctuations.---} Finally, it is interesting to relate the strong  cross-correlation amplitudes indicative for the presence of non-commuting operator states to the rare event statistics of current flow. To understand this point,  consider the  probability distribution of the currents $I_{1}$ and $I_2$, obtained from the generating function $\rho_{\tau}(z_1,z_2)$ in Eq.~\eqref{eq:fcs},
\begin{equation}\label{prob1}
P(I_1, I_2) = -\int \! \frac{d z_1}{2\pi z_1}\frac{dz_2}{2\pi z_2}\,\operatorname{Tr}\rho_{\tau}(z_1,z_2) z_1^{-I_1\tau}z_2^{-I_2\tau}.
\end{equation}
Focusing on the tails of the distribution, $I_\alpha \gg \bar I$, a straightforward saddle-point approximation stabilized by $\Gamma \tau \gg 1$~\cite{appendix} yields
\begin{equation}
    P(I_1, I_2) \stackrel{I_\alpha\gg \bar I}\simeq \prod_{\alpha} \left(\frac{\bar I}{I_\alpha}\right)^{I_\alpha \tau}.
\end{equation}
These tails decay exponentially, but much slower than for a Gaussian distribution. This reflects the fact that the simultaneous measurement of non-commuting operators triggers rare fluctuations  stronger than those caused by the superposition of uncorrelated fluctuations\cite{Wei2008}.

\emph{Conclusions.---} We have proposed an experimental diagnostic for MBS which,
much as a braiding protocol, probes the commutation relations of a Majorana algebra,
but should be experimentally feasible at drastically lower experimental effort. The
approach is based on monitoring the statistics of tunnel currents in response to changes of a few easily accessible system parameters, the gate-controlled tunneling contacts into the system.  The comparatively easy variability  of these parameters in one experimental run defines a structured pattern of quantitative  predictions, the `true Majorana case' being identified by a multitude of testable conditions (as opposed to just one signal in tunneling spectroscopy data.) We therefore believe, that the experiment would yield a definite fingerprint. Conceptually, it amounts
to a continuous weak measurement, a most direct approach to
probing the presence of non-commuting operators. Since the measurement outcome
qualitatively depends on the underlying operator algebra, the
recording of transport statistics as summarized in table~\ref{tab:cross_corr} would
represent compelling evidence for the presence of a Majorana qubit.
While we expect the qualitative distinction between MBS and ABS to display a high level of parameter tolerance, it will be rewarding to study non-equilibrium noise for microscopically more refined models. As with conventional quantum devices, the added information sitting in statistical fluctuations will provide a higher level of realistically accessible information on topological quantum wires than that provided by DC transport probes.

\begin{acknowledgments}
We thank D. Bagrets for useful discussions. This work has been funded by the Deutsche Forschungsgemeinschaft (DFG, German Research Foundation) under
Germany's Excellence Strategy – Cluster of Excellence Matter and Light for Quantum Computing
(ML4Q) EXC 2004/1 – 390534769. We also acknowledge DFG funding under Projektnummer 277101999 – TRR 183 (project C01 and Mercator program), and by the Danish National Research Foundation.
\end{acknowledgments}

\onecolumngrid\clearpage\setcounter{equation}{0}
\renewcommand\theequation{S\arabic{equation}}
\section{Supplement}\setcounter{page}{1}

\section{Derivation of Bloch Redfield equation with counting fields}

Here we provide details regarding the derivation of the evolution
equation {[}see Eqs. (6-8) in the main text{]}, which we use to model
the Majorana platforms and obtain the full counting statistics (FCS).
Throughout we will work in units with $k_{B}=e=\hbar=1$. It is convenient
to split the effective Hamiltonian according to 
\begin{equation}
H=H_{0}+H_{I}\label{eq:1}
\end{equation}
with one part being the lead Hamiltonian
\begin{equation}
H_{0}=\sum_{\alpha=0}^{2}\sum_{k}(\xi_{k}-V\delta_{\alpha0})c_{\alpha,k}^{\dagger}c_{\alpha,k}^{}\label{eq:2}
\end{equation}
with $\xi_{k}$ the single-particle energy. The interaction Hamiltonian is then given by 
\begin{equation}
H_{I}=H_{\mathrm{ref}}+\tilde{H}_{T},\label{eq:3}
\end{equation}
with $\tilde{H}_{T}$ {[}see Eq. (3) in the main text{]} and $H_{\mathrm{ref}}$
as defined in the main text. We want to work in the corresponding interaction
picture with $\rho_{I}(t)$ the interaction picture density matrix. Now we write the moment generating
function of the charge variables $Q_{\alpha}$ in lead $\alpha=1,2$
as 
\begin{equation}
\mathcal{Z}=\mathrm{Tr}\rho_{\tau}(z_{1},z_{2})\label{eq:4}
\end{equation}
with $\tau$ being the duration of measurement. Here we have defined a generalized
density matrix
\begin{equation}
\rho_{t}(z_{1},z_{2})=z_{1}^{Q_{1}(t)/2}z_{2}^{Q_{2}(t)/2}\rho_{I}(t)z_{1}^{Q_{1}(t)/2}z_{2}^{Q_{2}(t)/2}\label{eq:5}
\end{equation}
which obeys a generalized Liouville von Neumann equation 
\begin{equation}
\frac{\partial}{\partial t}\rho_{t}=-iH_{I}^{+}(t)\rho_{t}+i\rho_{t}H_{I}^{-}(t)\label{eq:6}
\end{equation}
where $H_{I}(t)$ is modified with counting parameters $z_{\alpha}$ to give
\begin{equation}
H_{I}^{\pm}=\sum_{\alpha>\alpha'}\frac{z_{\alpha}^{\pm\frac{1}{2}}}{z_{\alpha'}^{\pm\frac{1}{2}}}(t_{0,\alpha}\delta_{0\alpha'}+\mathcal{O}_{\alpha\alpha'})\sum_{k,k'}c_{\alpha,k}^{\dagger}c_{\alpha',k'}^{}+h.c.\label{eq:7}
\end{equation}
with $z_{0}\equiv1$. The operators $\mathcal{O}_{\alpha\alpha'}$
{[}see Eq. (3) in the main text{]} obey $\mathcal{O}_{\alpha\alpha'}=\mathcal{O}_{\alpha'\alpha}^{\dagger}$.
By applying a standard Born-Markov approximation we obtain
\begin{eqnarray}
\dot{\rho}_{q} & = & \intop_{0}^{\infty}ds\mathrm{Tr}_{\mathrm{L}}\left(H_{I}^{+}(t)\rho_t H_{I}^{-}(t-s)+H_{I}^{+}(t-s)\rho_t H_{I}^{-}(t)\right.\nonumber \\
 &  & \left.-H_{I}^{+}(t)H_{I}^{+}(t-s)\rho_t-\rho_t H_{I}^{-}(t-s)H_{I}^{-}(t)\right),\label{eq:10}
\end{eqnarray}
with $\rho_{q}=\rho_{q}(t)$ and
\begin{equation}
\rho_{t}\simeq\rho_{q}(t)\otimes\rho_{\mathrm{L}}\label{eq:8}
\end{equation}
where $\rho_{\mathrm{L}}$ is the lead density matrix. This is justified for sufficiently weak coupling between the Majorana qubit and the reservoirs. Here the symbol $\mathrm{Tr}_{\mathrm{L}}$ stands for the trace over the leads. Further steps in the standard derivation of (8) involved the replacement of $\rho_{s}$ under
the integral by $\rho_{t}$ as well as to extend the time integration
to infinity. Now we relabel
\begin{equation}
\rho_{q}\rightarrow\rho\label{eq:11}
\end{equation}
and write the equation in the form 
\begin{equation}
\frac{\partial}{\partial t}\rho=\intop_{0}^{\infty}ds\left(A_{1}(s)+A_{2}(s)+B(s)\right).\label{eq:12}
\end{equation}
Let us introduce the shorthand notation
\begin{equation}
Q_{\alpha\alpha',t}=\sum_{k,k'}c_{\alpha,k}^{\dagger}(t)c_{\alpha',k'}^{}(t)\label{eq:13}
\end{equation}
and $\left\langle \ldots\right\rangle =\mathrm{Tr}_{\mathrm{L}}\left(\rho_{\mathrm{L}}\ldots\right).$
Then, with $z_{\alpha\alpha'}=z_{\alpha}/z_{\alpha'}$ the quantities
$B(s)$ and $A_{\alpha}(s)$ are given by
\begin{eqnarray}
B & = & \left(z_{21}\mathcal{O}_{12}^{\dagger}\rho\mathcal{O}_{12}^{}-\mathcal{O}_{12}^{}\mathcal{O}_{12}^{\dagger}\rho\right)\left\langle Q_{12,t}^{}Q_{12,t-s}^{\dagger}\right\rangle \label{eq:14} \\
 &  & +\left(z_{21}\mathcal{O}_{12}^{\dagger}\rho\mathcal{O}_{12}^{}-\rho\mathcal{O}_{12}^{}\mathcal{O}_{12}^{\dagger}\right)\left\langle Q_{12,t-s}^{}Q_{12,t}^{\dagger}\right\rangle \nonumber \\
 &  & +\left(z_{12}\mathcal{O}_{12}^{}\rho\mathcal{O}_{12}^{\dagger}-\rho\mathcal{O}_{12}^{\dagger}\mathcal{O}_{12}^{}\right)\left\langle Q_{12,t-s}^{\dagger}Q_{12,t}^{}\right\rangle \nonumber \\
 &  & +\left(z_{12}\mathcal{O}_{12}^{}\rho\mathcal{O}_{12}^{\dagger}-\mathcal{O}_{12}^{\dagger}\mathcal{O}_{12}^{}\rho\right)\left\langle Q_{12,t}^{\dagger}Q_{12,t-s}^{}\right\rangle \nonumber
\end{eqnarray}
and
\begin{eqnarray}
A_{\alpha} & = & t_{0,\alpha}[\mathcal{O}_{\alpha0}^{\dagger},\rho]\left\langle Q_{\alpha0,t-s}^{}Q_{\alpha0,t}^{\dagger}-Q_{\alpha0,t}^{\dagger}Q_{\alpha0,t-s}^{}\right\rangle \nonumber \\
 &  & +t_{0,\alpha}^{\ast}[\mathcal{O}_{\alpha0},\rho]\left\langle Q_{\alpha0,t-s}^{\dagger}Q_{\alpha0,t}^{}-Q_{\alpha0,t}^{}Q_{\alpha0,t-s}^{\dagger}\right\rangle \nonumber \\
 &  & +\left(z_{0\alpha}\mathcal{O}_{\alpha0}^{\dagger}\rho\mathcal{O}_{\alpha0}^{}-\mathcal{O}_{\alpha0}^{}\mathcal{O}_{\alpha0}^{\dagger}\rho\right)\left\langle Q_{\alpha0,t}^{}Q_{\alpha0,t-s}^{\dagger}\right\rangle \nonumber \\
 &  & +\left(z_{0\alpha}\mathcal{O}_{\alpha0}^{\dagger}\rho\mathcal{O}_{\alpha0}^{}-\rho\mathcal{O}_{\alpha0}^{}\mathcal{O}_{\alpha0}^{\dagger}\right)\left\langle Q_{\alpha0,t-s}^{}Q_{\alpha0,t}^{\dagger}\right\rangle \nonumber \\
 &  & +\left(z_{\alpha0}\mathcal{O}_{\alpha0}^{}\rho\mathcal{O}_{\alpha0}^{\dagger}-\rho\mathcal{O}_{\alpha0}^{\dagger}\mathcal{O}_{\alpha0}^{}\right)\left\langle \mathcal{O}_{\alpha0,t-s}^{\dagger}Q_{\alpha0,t}^{}\right\rangle \nonumber \\
 &  & +\left(z_{\alpha0}\mathcal{O}_{\alpha0}^{}\rho\mathcal{O}_{\alpha0}^{\dagger}-\mathcal{O}_{\alpha0}^{\dagger}\mathcal{O}_{\alpha0}^{}\rho\right)\left\langle \mathcal{O}_{\alpha0,t}^{\dagger}Q_{\alpha0,t-s}^{}\right\rangle \nonumber \\
 &  & +(z_{0\alpha}-1)\left[t_{0,\alpha}^{\ast}\rho\mathcal{O}_{\alpha0}^{}+t_{0,\alpha}\mathcal{O}_{\alpha0}^{\dagger}\rho+|t_{0,\alpha}|^{2}\rho\right]\nonumber \\
 &  & \qquad\times\left\langle Q_{\alpha0,t-s}^{}Q_{\alpha0,t}^{\dagger}+Q_{\alpha0,t}^{}Q_{\alpha0,t-s}^{\dagger}\right\rangle \nonumber \\
 &  & +(z_{\alpha0}-1)\left[t_{0,\alpha}^{\ast}\mathcal{O}_{\alpha0}^{}\rho+t_{0,\alpha}\rho\mathcal{O}_{\alpha0}^{\dagger}+|t_{0,\alpha}|^{2}\rho\right]\nonumber \\
 &  & \qquad\times\left\langle Q_{\alpha0,t}^{\dagger}Q_{\alpha0,t-s}^{}+Q_{\alpha0,t-s}^{\dagger}Q_{\alpha0,t}^{}\right\rangle .\label{eq:15}
\end{eqnarray}
The next step is to trace out the fermionic reservoirs. We approximate the latter to be in thermal equilibrium,
\begin{equation}
\rho_{\mathrm{L}}\sim e^{-\frac{1}{T} H_{\mathrm{0}}}, \label{eq:16}
\end{equation}
since the coupling due to $H_{\mathrm{ref}}$ and $\tilde{H}_{T}$ is weak. 
Exemplarily this yields relations like
\begin{equation}
\intop_{0}^{\infty}ds\left\langle Q_{\alpha0,t-s}^{\dagger}Q_{\alpha0,t}^{}-Q_{\alpha0,t}^{}Q_{\alpha0,t-s}^{\dagger}\right\rangle =\nu^{2}(i\Lambda+\pi V),\label{eq:17}
\end{equation}
where $\Lambda$ is the bandwidth of the leads and $\nu$ the density
of states. Analogously we obtain for instance
\begin{equation}
\intop_{0}^{\infty}ds\left\langle Q_{\alpha0,t-s}^{}Q_{\alpha0,t}^{\dagger}+Q_{\alpha0,t}^{}Q_{\alpha0,t-s}^{\dagger}\right\rangle =2\pi\nu^{2}Vn_{B}(V),\label{eq:18}
\end{equation}
where $n_B(V)$ is the Bose function which is negligible in the limit of interest $T\ll V$. The terminals $\alpha=1,2$ are at the same potential such that thermal cotunneling processes are relevant here, 
\begin{equation}
\intop_{0}^{\infty}ds\left\langle Q_{12,t-s}^{}Q_{12,t}^{\dagger}+Q_{12,t}^{}Q_{12,t-s}^{\dagger}\right\rangle =2\pi\nu^{2}T.\label{eq:19}
\end{equation}
By evaluating all terms in Eq. (11) along these lines, and using the approximations $\Lambda-i\pi V\simeq \Lambda$ and $1+\ln(\Lambda/2V) \simeq \ln(\Lambda/2V)$ in $H_q$, we arrive at Eqs. (6-8) as quoted in the main text. The necessary formalism to treat the problem of weak continuous measurement of non-commuting variables has been discussed in Ref. [101, 103, 104] of the Letter where similar evolution equations have been studied.

\section{The case of true Majoranas}

In this section we provide details regarding our analysis of true
(single) Majoranas ($N_{\alpha}=1$). To this end we set $\mathcal{O}_{\alpha\alpha'}=t_{1}(i\gamma_{\alpha}\gamma_{\alpha'})$ and $t_{0,\alpha}=t_{0}$. Furthermore, we rewrite $\rho_{t}=e^{\theta t}\tilde{\rho}_{t}$ with 
\begin{eqnarray}
\theta & = & 2\pi\nu^{2}(|t_{0}|^{2}+|t_{1}|^{2})V\sum_{\alpha=1}^{2}(z_{\alpha}-1)\nonumber \\
 &  & +2\pi\nu^{2}|t_{1}|^{2}T(z_{1}/z_{2}+z_{2}/z_{1}-2).\label{eq:63}
\end{eqnarray}
Then $\tilde{\rho}_{t}\equiv\tilde{\rho}_{t}(z_{1},z_{2})$ obeys
the evolution equation 
\begin{eqnarray}
\dot{\tilde{\rho}}_{t} & = & -i[\tilde H_q,\tilde{\rho}_{t}]+\sum_{\alpha>\alpha'}\frac{\varGamma_{\alpha\alpha'}}{2}\left((i\gamma_{\alpha}\gamma_{\alpha'})\tilde{\rho}_{t}(i\gamma_{\alpha}\gamma_{\alpha'})-\tilde{\rho}_{t}\right)\nonumber \\
 &  & +2\pi V\sum_{\alpha=1}^{2}(z_{\alpha}-1)\left(t_{0}^{\ast}t_{1}\sigma_{\alpha}\tilde{\rho}_{t}+t_{0}t_{1}^{*}\tilde{\rho}_{t}\sigma_{\alpha}\right).\label{eq:64}
\end{eqnarray}
Here we defined $\varGamma_{\alpha0}=\varGamma z_{\alpha}$ with $\varGamma\equiv4\pi V\nu^{2}|t_{1}|^{2}$
and $\varGamma_{21}\equiv4\pi\nu^{2}T|t_{1}|^{2}(z_{1}/z_{2}+z_{2}/z_{1})$
as well as the Hamiltonian 
\begin{equation}
\tilde H_q=-2\nu^{2}(\Lambda\mathrm{Re}(t_{0}^{\ast}t_{1})+\pi V\mathrm{Im}(t_{0}^{\ast}t_{1}))\sum_{\alpha=1}^{2}\sigma_{\alpha}.\label{eq:H_q=00003D00003D-1}
\end{equation}
To solve (\ref{eq:64}) we parametrize the density matrix as $\rho_{t}=\sum_{\mu=0}^{3}\rho_{\mu,t}\sigma_{\mu}$
with $\sigma_{1}=i\gamma_{1}\gamma_{0}$, $\sigma_{2}=i\gamma_{2}\gamma_{0}$,
$\sigma_{3}=i\gamma_{2}\gamma_{1}$ and $\sigma_{0}=\mathbb{I}$.
We obtain the first order system
\begin{equation}
\frac{\partial}{\partial t}\tilde{\rho}_{\mu,t}(z_{1},z_{2})=\sum_{\mu=0}^{3}M_{\mu\nu}\tilde{\rho}_{\nu,t}(z_{1},z_{2}).\label{eq:rho=00003D00003DMrho-1}
\end{equation}
The matrix $M$ of coefficients reads 
\begin{equation}
M=\left(\begin{array}{cccc}
0 & a_{1} & a_{2} & 0\\
a_{1} & -\varGamma_{20}-\varGamma_{21} & 0 & h_{2}\\
a_{2} & 0 & -\varGamma_{10}-\varGamma_{21} & -h_{1}\\
0 & -h_{2} & h_{1} & -\varGamma_{10}-\varGamma_{20}
\end{array}\right)\label{eq:M=00003D00003D}
\end{equation}
with 
\begin{equation}
a_{\alpha}=4\pi V\nu^{2}\mathrm{Re}(t_{0}^{\ast}t_{1})(z_{\alpha}-1),\label{eq:a=00003D00003D}
\end{equation}
\begin{equation}
h_{\alpha}=-4\nu^{2}\Lambda\mathrm{Re}(t_{0}^{\ast}t_{1})-4\pi\nu^{2}V\mathrm{Im}(t_{0}^{\ast}t_{1})z_{\alpha}.\label{eq:h=00003D00003D}
\end{equation}
The solution is given by a matrix exponential 
\begin{equation}
\rho_{\mu,\tau}(z_{1},z_{2})=e^{\theta\tau}\sum_{\nu=0}^{3}\exp(\tau M)_{\mu\nu}\rho_{\nu,0},\label{eq:solution as matrix exponential}
\end{equation}
where $\rho_{0}$ is the initial reduced density matrix. In the long
time limit $\varGamma\tau\gg1$ the cumulant generating function reads
\begin{equation}
\ln\mathcal{Z}=\tau\theta(z_{1},z_{2})+\tau\lambda_{0}(z_{1},z_{2}).\label{eq:lnZ=00003D00003Dlamda0}
\end{equation}
Here $\lambda_{0}(z_{1},z_{2})$ is the (unique) solution of the characteristic
polynomial of $M$ which satisfies $\lambda_{0}(1,1)=0$. The other
eigenvalues $\lambda_{i=1,2,3}$ of $M$ have a negative real part
at $(z_{1},z_{2})=(1,1)$. Expansion of (\ref{eq:lnZ=00003D00003Dlamda0})
in the bias yields Eq. (9) in the main text.

\subsection*{Large fluctuations}

Here we provide further details on our analysis of large fluctuations.
In the saddle point approximation valid for $\varGamma\tau\gg1$ we
obtain
\begin{equation}
\ln P(I_{1},I_{2})\simeq\min_{\mu_{1},\mu_{2}}\left(\ln\mathrm{Tr}\rho_{\tau}(e^{-\mu_{1}},e^{-\mu_{2}})+\sum_{\alpha}\mu_{\alpha}I_{\alpha}\tau\right).\label{eq:LF}
\end{equation}
For rare events, $I_{\alpha}\gg\bar{I}$, we find minima located at
$\mu_{\alpha}\sim\ln\left(\bar{I}/I_{\alpha}\right)$ resulting in
the factorized probability distribution presented in Eq. (13) in the
main text.

\subsection*{Timescale of projective measurement}

The timescale of projective measurement referred to in the discussion
of quasiparticle poisoning in the main text can be estimated by the time
$\tau_{\mathrm{proj}}\simeq S/(\Delta I)^{2}$ it takes to reach a
signal to noise ratio of order unity. From Eq. (4) in the main text,
we obtain the difference in currents to be $\text{\ensuremath{\Delta}}I=4g_{\mathrm{i}}V$.
The noise is given by the Schottky formula $S=\left\langle I_{2}\right\rangle $.
This leads to the timescale
\begin{equation}
\tau_{\mathrm{proj}}\simeq\frac{g_{0}+g_{1}}{16g_{\mathrm{i}}^{2}V}.\label{eq:proj}
\end{equation}

\section{the case of Andreev bound states}

In this section we provide details regarding our analysis of fermionic
zero-energy Andreev bound states (ABS). We model the latter as a pair
of two nonoverlapping Majorana states which are both coupled to the
respective terminal. In the following we adopt the shortened notation
$t_{1,\alpha\alpha'}^{ij}\equiv t_{\alpha\alpha'}^{ij}$.

\subsection*{FCS for source lead coupled to ABS}

For $N_{0}=2$, $N_{1,2}=1$ and in the absence of the reference arms
we find that $\tau^{-1}\ln\mathcal{Z}$ is given by the eigenvalue
of $\tilde{\theta}\mathbb{I}+\tilde{M}$ which vanishes at $(z_{1},z_{2})=(1,1)$.
Here we have defined
\begin{equation}
\tilde{\theta}=2\pi\nu^{2}V\sum_{\alpha=1}^{2}\sum_{i=1}^{2}|t_{\alpha0}^{1i}|^{2}(z_{\alpha}-1)\label{eq:theta 121}
\end{equation}
and the matrix
\begin{equation}
\tilde{M}=4\pi\nu^{2}V\sum_{\alpha=1}^{2}\left(\begin{array}{cc}
0 & -b_{\alpha}(z_{\alpha}-1)\\
b_{\alpha}(z_{\alpha}+1) & -(|t_{\alpha0}^{11}|^{2}+|t_{\alpha0}^{12}|^{2})z_{\alpha}
\end{array}\right)\label{31-1}
\end{equation}
with $b_{\alpha}=\mathrm{Im}(t_{\alpha0}^{11}(t_{\alpha0}^{12})^{\ast})$.
For $|t_{1,10}^{ij}|=|t_{1,20}^{ij}|$ one obtains
\begin{equation}
S_{12}=2\pi\nu^{2}V\left[\frac{\sum_{\alpha=1}^{2}b_{\alpha}^{2}}{|t_{10}^{11}|^{2}+|t_{10}^{12}|^{2}}-\frac{2b_{1}b_{2}\left(b_{1}+b_{2}\right)^{2}}{\left(|t_{10}^{11}|^{2}+|t_{10}^{12}|^{2}\right)^{3}}\right]\label{eq:66-1-1}
\end{equation}
at zero temperature. If we further assume for simplicity $\mathrm{Im}((t_{10}^{11})^{\ast}t_{10}^{12})=-\mathrm{Im}((t_{20}^{11})^{\ast}t_{20}^{12})$
we obtain Eq. (11) in the main text. Whenever the condition $N_{0}=2$
is fulfilled the Fano factor will satisfy generically $F=\mathcal{O}(1)$.

\subsection*{FCS for source coupled to MBS and drains to ABS}

Now we focus on the case $N_{1,2}=2$, $N_{0}=1$ in the absence of
the reference arms. Consider
\begin{equation}
\theta'=2\pi\nu^{2}V\sum_{\alpha=1}^{2}\sum_{i=1}^{2}|t_{\alpha0}^{i1}|^{2}(z_{\alpha}-1)\label{eq:56}
\end{equation}
and the matrix
\begin{equation}
M'=\left(\begin{array}{cccc}
0 & c_{1,-} & c_{2,-} & 0\\
-c_{1,+} & -d_{1} & 0 & c_{2,-}\\
-c_{2,+} & 0 & -d_{2} & c_{1,-}\\
0 & -c_{2,+} & -c_{1,+} & -d_{1}-d_{2}
\end{array}\right)\label{eq:67}
\end{equation}
with 
\begin{equation}
c_{\alpha,\pm}=4\pi\nu^{2}V\mathrm{Im}(t_{\alpha0}^{11}(t_{\alpha0}^{21})^{\ast})(z_{\alpha}\pm1),\label{eq:c=00003D00003D-1}
\end{equation}
\begin{equation}
d_{\alpha}=4\pi\nu^{2}V\sum_{i=1}^{2}|t_{\alpha0}^{i1}|^{2}z_{\alpha}.\label{eq:h=00003D00003D-1-1}
\end{equation}
Now $\tau^{-1}\ln\mathcal{Z}$ is given by the eigenvalue of $\theta'\mathbb{I}+M'$
which vanishes at $(z_{1},z_{2})=(1,1)$. At zero temperature we obtain
$S_{12}=0$ to the leading order. The implication is that a Fano factor $F=\mathcal{O}(1)$
without reference arms requires $N_{0}=2$ as stated in
Table I in the main text. 

\begin{thebibliography}{99}
\bibitem{footnote}
The result~\eqref{eq:Ftrue} displays a high degree of robustness with respect to finite coupling terms $\sim i\gamma_1\gamma_0$ and $\sim i\gamma_2\gamma_0$ in the effective Hamiltonian $H_q$. We find a slightly higher degree of sensitivity with respect to the term $\varepsilon_{12}i\gamma_1\gamma_2$. However, $F$ stays of order unity as long as $\varepsilon_{12}^2 \lesssim \Lambda^2 [\nu^2 \operatorname{Re}(t_0^*t_1)]^2$ is fulfilled. We note that in our proposed geometries $\gamma_1$ and $\gamma_2$ are hosted by different topological wires such that we expect the corresponding hybridization $|\varepsilon_{12}|$ to be sufficiently small.
\end{thebibliography}
\end{document}